\documentclass{aa}
\usepackage{graphicx}
\usepackage{txfonts}
\usepackage{natbib}
\bibpunct{(}{)}{;}{a}{}{,}

\begin{document}

\title{Detection of X-rays from the jet-driving Symbiotic Star MWC 560}

\author{Matthias Stute\inst{1} \and Raghvendra Sahai\inst{2}}

\offprints{Matthias Stute, \\ \email{mstute@phys.uoa.gr}}

\institute{IASA and Section of Astrophysics, Astronomy and Mechanics,
Department of Physics, University of Athens, Panepistimiopolis, 
15784 Zografos, Athens, Greece
\and
Jet Propulsion Laboratory, California Institute of Technology, 
4800 Oak Grove Drive, Pasadena, CA 91109, USA}

\date{Received 17 October 2008; accepted 01 February 2009}

\abstract
{}
{We report the detection of X-ray emission from the jet-driving symbiotic 
star MWC\,560.}
{We observed MWC\,560 with XMM-Newton for 36 ks. We fitted the spectra from 
the EPIC pn, MOS1 and MOS2 instruments with XSPEC and examined the light curves
with the package XRONOS.}
{The spectrum can be fitted with a highly absorbed hard X-ray 
component from an optically-thin hot plasma, a Gaussian emission line with an
energy of 6.1 keV and a less absorbed soft thermal component. The best fit is 
obtained with a model in which the hot component is produced by optically thin 
thermal emission from an isobaric cooling flow with a maximum temperature of 61 
keV, which might be created inside an optically-thin boundary layer on the 
surface of the accreting with dwarf. The derived parameters of the hard 
component detected in MWC\,560 are in good agreement with similar objects as 
CH\,Cyg, SS7317, RT\,Cru and T\,CrB, which all form a new sub-class of 
symbiotic stars emitting hard X-rays. Our previous numerical simulations of the 
jet in MWC\,560 showed that it should produce detectable soft X-ray emission. 
We infer a temperature of 0.17 keV for the observed soft component, i.e. less 
than expected from our models. The total soft X-ray flux (i.e. at $<$ 3 keV) is 
more than a factor 100 less than predicted for the propagating jet soon after 
its birth ($<0.3$ yr), but consistent with the value expected due its decrease 
with age. The ROSAT upper limit is also consistent with such a decrease. We 
find aperiodic or quasi-periodic variability on timescales of minutes and 
hours, but no periodic rapid variability.
}
{All results are consistent with an accreting white dwarf powering the X-ray 
emission and the existence of an optically-thin boundary layer around it.
}


\maketitle

\section{Introduction}

Highly collimated fast outflows or jets are common in many astrophysical 
objects of different sizes and masses: active galactic nuclei (AGN), X-ray 
binaries (XRBs), young stellar objects (YSO), pre-planetary nebulae (PPN), 
supersoft X-ray sources and symbiotic stars. However, there are still open 
questions concerning the formation of jets.

In current jet formation models, accretion disks and magnetic fields play a 
key role. In their analytical model, \citet{BlP82} invoked magneto-centrifugal 
acceleration along magnetic field lines threading an accretion disk -- the 
poloidal magnetic field component brakes the disk matter and accelerates it 
into the jet and the toroidal component collimates the flow. 

Symbiotic stars are ideal test beds for studying the formation and propagation 
of jets in binary systems where an accretion disk is believed to exist around 
a white dwarf (WD). Symbiotic stars are interacting binaries with orbital 
periods in the range of years; the cool component is a red giant (RG), the hot 
component a WD (with Teff $\sim$ 50000 -- 200000 K). Both stars show mass loss 
through supersonic winds. Wind material from the RG is captured by the WD to 
form an accretion disk. Prominent jets have been observed in symbiotic stars, 
in stark contrast to other binary systems with WD companions, like cataclysmic 
variables which also harbor accretion disks, yet show no jets.

About 200 symbiotic stars are known \citep[e.g.][]{BMM00}, but jets have been 
detected at different wavelengths only in 10 out of them \citep{BSK04}. The 
most famous systems are R\,Aquarii, CH\,Cygni and MWC\,560. X-ray observations 
provide a direct probe of the two most important components of jet-driving 
systems: the bow and internal shocks of the jet and the central parts of the 
jet engine, where gas is being accreted for powering the jet. Up to now, R\,Aqr 
and CH\,Cyg are the only two jets of symbiotic stars which are detected in 
X-rays. 

R\,Aquarii, with a distance of about 200 pc, is one of the nearest symbiotic 
stars and a well known jet source. The jet has been extensively observed in 
the optical and at radio wavelengths \citep[e.g.][]{SoU85,PaH94,HMK85,HLD85}.
It shows a rich morphology with hints of pulsed ejection. Furthermore, R\,Aqr is
the first jet in a symbiotic system, which was detected in X-rays 
\citep{VPF87,HSS98,KPL01}. The spectra of the jets are consistent with a soft 
component with $k\,T \sim$ 0.25 keV. The central source shows a supersoft 
blackbody emission with $k\,T \sim$ 0.18 keV and a Fe K$\alpha$ line at 6.4 
keV which suggests the presence of a hard source near the hot star. Recently, 
\citet{KAK07} reported on five years of observations with Chandra and were able
to measure the proper motion of knots in the NE jet of about 600 km s$^{-1}$. 
\citet{NDK07} investigated the X-ray emission from the inner 500 AU of this 
system and were able to observe a hard thermal component with a temperature of 
6.8 keV.

In 1984/85, CH\,Cygni showed a strong radio outburst, during
which a double-sided jet with multiple components was ejected \citep{TSM86}.
This event allowed an accurate measurement of the jet velocity near 1500 km 
s$^{-1}$. In HST observations \citep{EBS02}, arcs can be detected that also 
could be produced by episodic ejection events. X-ray emission was first 
detected by EXOSAT \citep{LeT87}, and subsequent ASCA observations 
revealed a complex X-ray spectrum with two soft components ($k\,T =$ 0.2 and 
0.7 keV) associated with the jet, an absorbed hard component (7.3 keV) and a 
Fe K$\alpha$ line \citep{EIM98}. Analysis of archival Chandra ACIS data by 
\citet{GaS04} revealed faint extended emission to the south, aligned with the 
optical and radio jets seen in HST and VLA observations. An apparent decline of 
the hard X-ray component has been observed with the US-Japanese X-ray satellite 
Suzaku by \citet{MIK07}. Recently, \citet{KCR07} reported the detection of 
multiple spatial components, including an arc, in the archival Chandra images.

In both objects, hints of variability of the hard X-ray component were found 
with periods of 1734s in R\,Aqr \citep{NDK07} and 130s in\,CH Cyg \citep{EIM98},
however, not with high confidence levels. If present, these variabilities can 
be interpreted as the Keplerian period at the magnetospheric radius, where the 
accretion disk is truncated by a strong magnetic field of the WD.

While these two objects are seen at high inclinations, in MWC\,560 (or V694 
Mon) the jet axis is practically parallel to the line of sight 
\citep{TKG90,SKC01}. This special orientation allows one to observe the 
outflowing gas as line absorption in the source 
spectrum. With such observations the radial velocity and the column density of 
the outflowing jet gas close to the source has been investigated in great 
detail. In particular the acceleration and evolution of individual outflow 
components and jet pulses has been probed with spectroscopic monitoring 
programs, as described in \citet{SKC01}. Using this optical data, 
sophisticated numerical models of this pulsed propagating jet have been 
developed \citep{SCS05,Stu06}. A number of hydrodynamical simulations (with 
and without cooling) were made in which the jet density and velocity during 
the pulses were varied. The basic model absorption line profiles in MWC\,560 
as well as the mean velocity and velocity-width are in good agreement with the 
observations. The evolution of the time-varying high velocity absorption 
line-components is also well modeled. These models not only fit the MWC\,560 
data, but are also able to explain properties of jets in other symbiotic 
systems such as the observed velocity and temperature of the CH\,Cyg jet. Until 
recently, no hydrodynamical models existed for explaining the X-ray emission 
from symbiotic stars, thus as a first step we have therefore used our existing 
simulations, which fit MWC\,560, for understanding the observed X-ray emission 
properties of CH\,Cyg and R\,Aqr and the expected properties of MWC\,560 
\citep{StS07}. In our models, the jet emits two soft one-temperature components,
similar to those seen in CH\,Cyg. 

MWC\,560 has not been detected in X-rays so far -- only ROSAT observations have
been available, with an upper limit of $7\times 10^{-4}$ counts s$^{-1}$ 
\citep{MWJ97} corresponding to $8\times10^{-15}$ ergs\,cm$^{-2}$\,s$^{-1}$ 
(PIMMS). Here, finally, we report on the analysis of our recent {\em XMM-Newton}
observations of MWC\,560 and on the detection of X-ray emission from this 
object. 

The remainder is organized as follows: in \S \ref{sec_obs}, we show details of 
the observations and the analysis of the data. After that we describe the 
results in \S \ref{sec_res}. We end with a discussion and conclusions. 

\section{Observation and Analysis} \label{sec_obs}

The {\em XMM-Newton} X-ray observatory observed the field of MWC\,560 for 
33 ks in 2007 on September 27, from 02:17 UT to 12:22 UT (observation ID: 
0501350101). The data were collected with the EPIC instrument, which consists 
of two MOS \citep{TAA01} and one pn \citep{SBD01} cameras sensitive to photons 
with energies between 0.15 and 15 keV. The EPIC pn as well as EPIC MOS were 
operated in Full Window mode, i.e. with a field of view of 30'. Both the pn and 
MOS mounted the medium thickness filter. Simultaneously, the observations 
provided optical imaging from the Optical Monitor \citep[OM;][]{MBM01}. Because 
of a low count rate, the Reflection Grating Spectrograph data were not useful. 
Further details are given in Table \ref{tbl_obs}.

All the data reduction was performed using the Science Analysis Software 
({SAS}) software package\footnote{See \mbox{http://xmm.vilspa.esa.es/}.}
version 7.1. The raw observation data files were processed using standard 
pipeline tasks ({epproc} for pn, {emproc} for MOS data).
We selected events with pattern 0--4 (only single and double events) for the pn
and pattern 0--12 for the MOS, respectively, and applied the filter 
{FLAG=0}.

\begin{table*}
\caption{Observations on September 27, 2007}
\label{tbl_obs}
\centering
\begin{tabular}{lllll} 
\hline\hline
Instrument & Filter & Duration (s) & UT Start Time & UT Stop Time \\
\hline
PN   & Medium & 34457 & 02:40:19 & 12:14:36 \\
MOS1 & Medium & 36019 & 02:18:22 & 12:18:41 \\
MOS2 & Medium & 36026 & 02:18:21 & 12:18:47 \\
\hline
\end{tabular}
\end{table*}

\subsection{Spectral analysis}

\begin{figure}
  \centering
  \includegraphics[width=\columnwidth]{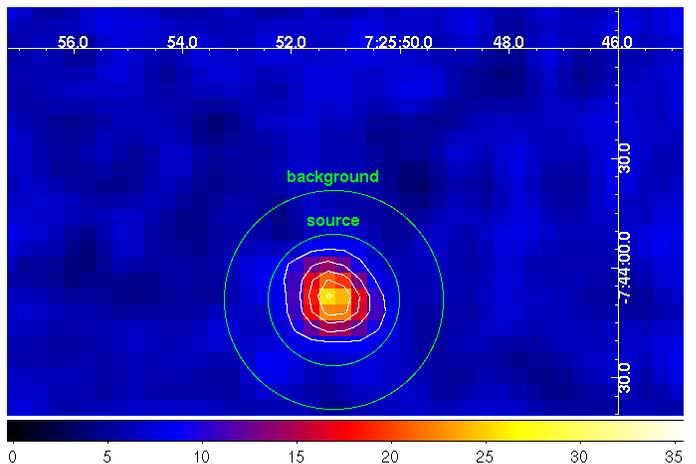}
  \caption{{\em XMM-Newton} EPIC pn image (detail) of the field of MWC\,560 in 
    the 0.2--15 keV range. Also shown are the extraction regions for the source 
    and the background events. Contour levels are 9, 12.6, 16.3 and 20 counts.}
  \label{image_PN}
\end{figure}

The source spectra were accumulated from a circular region (360 pixels 
radius, 18'') centered on MWC\,560, using the pn and also the MOS1 and MOS2 
detectors. The background spectra were extracted from a source-free region of 
the same chip taken with an annulus centered on the source around the former 
region with an outer radius of 600 pixels (30'', Fig. \ref{image_PN}).

Spectral redistribution matrices and ancillary response files were generated 
using the {SAS} scripts {rmfgen} and {arfgen}, and 
spectra grouped with a minimum of 25 counts per energy bin were fed into the 
spectral fitting package {XSPEC}\footnote{See 
\mbox{http://heasarc.gsfc.nasa.gov/docs/xanadu/xspec/}.} version 12.4.0. 
Spectral channels having energies below 0.1 keV and bad channels were ignored.

\subsection{Timing analysis}

We searched for variabilities in the X-ray emission from MWC\,560 using the pn
detector. Source photons were again selected from a circular region centered on 
MWC\,560 with a radius of 360 pixels (18'') and background photons from an 
annulus around the former region with an outer radius of 600 pixels (30'', Fig. 
\ref{image_PN}). Photon arrival times were converted to the solar system 
barycentre using the SAS task {barycen}. 

We searched the data for pulsations over a wide period range using 
{efsearch} and {powspec} within the package 
{XRONOS}\footnote{See 
\mbox{http://heasarc.gsfc.nasa.gov/docs/xanadu/xronos/xronos.html}.} version 
5.21. 

\section{Results} \label{sec_res}

\subsection{Spectra}

We find 624 counts in our source annuli in the EPIC pn detector, 150 counts in 
MOS1 and 137 counts in MOS2, respectively, all in the range of 0.2 -- 15 keV. 
In the soft range 0.2 -- 2 keV, where we expect emission from the jet, we 
detect 120, 24 and 26 counts, respectively.

First, we fit simultaneously the spectra from EPIC pn, MOS1 and MOS2 to a number
of different one-component models including a Raymond-Smith plasma \citep{RS77},
{MEKAL} \citep{MGO85,MLO86,LOG95}, and {APEC} \citep{SBL01}, all 
corrected for interstellar absorption. The abundances used are those of 
\citet{AnG89}. The fits are already very good with reduced $\chi^2$ values 
between 1.185 and 1.156. The model fits show a highly absorbed hard 
($k\,T \sim$ 11.3--13.2 keV) thermal component. The value for the column 
density of absorbing material is about $2.8\times10^{23}$ cm$^{-2}$. 
The null hypothesis probability, $P_{\rm nh}$ hereafter, is between 23 and 26 
\%.

Since in the residuals a peak between 6 and 7 keV is visible, we fit the  
spectra with two components as a next step, with the same {APEC} model 
with frozen parameters and an additional Gaussian representing an emission line.
The fit improve significantly, the reduced $\chi^2$ drops to 0.722 and
$P_{\rm nh}$ increases to 84 \%. The center of the line is found to be at 6.08 
keV with a line width of 0.23 keV. However, the fit to the \mbox{norm (GS)} 
parameter (Table \ref{Tbl_fits}), $2.74^{+1.63}_{-2.74}\times10^{-6}$, shows that
the data are also consistent with no line at all.

\begin{figure}
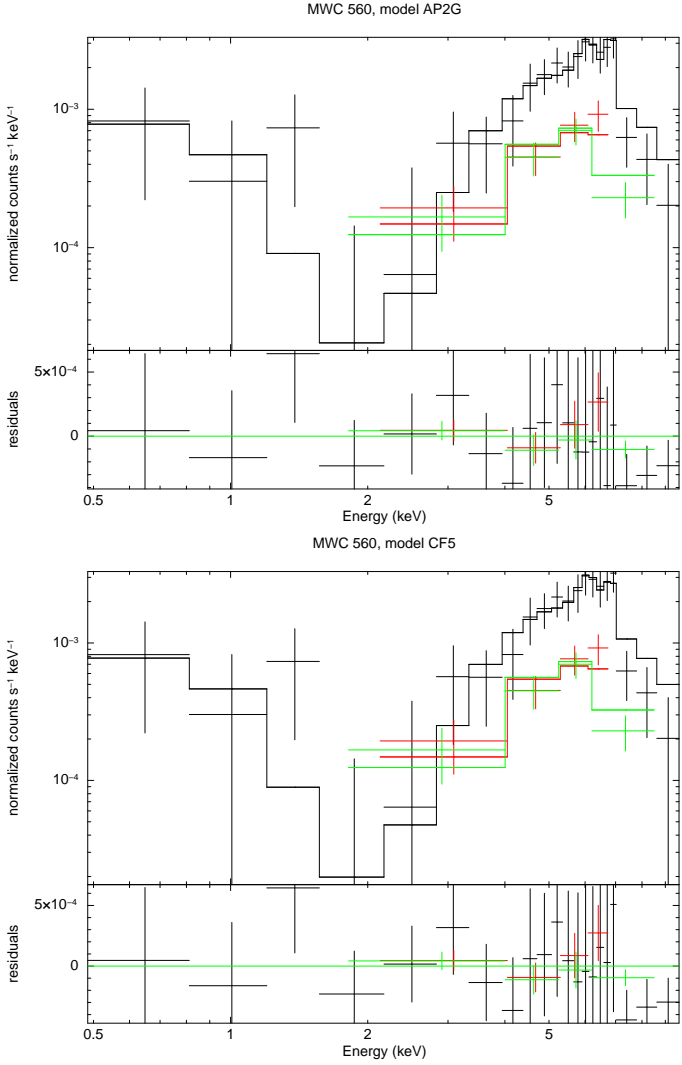

  \centering
  \includegraphics[angle=-90,width=\columnwidth]{fig2a.eps}
  \includegraphics[angle=-90,width=\columnwidth]{fig2b.eps}
  \caption{Observed spectrum of MWC\,560 together with several models: 
    our best fit models with an absorbed soft {APEC} model and another 
    absorbed hard {APEC} model plus a Gaussian (top) and an optically 
    thin thermal emission from an isobaric cooling flow with a maximum 
    temperature plus a Gaussian, both absorbed, and an additional absorbed soft 
    thermal component (bottom); also shown are the residuals for each fit. The 
    colors correspond to different instruments: PN (black), MOS1 (red), MOS2 
    (green)}
  \label{Fig_fit}
\end{figure}

Finally we add one absorbed soft thermal component and find a temperature of 
$k\,T \sim 0.18$ and an absorbing column density of $6.1\times10^{21}$ 
cm$^{-2}$. The reduced $\chi^2$ drops to 0.628 and $P_{\rm nh}$ is now 92 \%. 
The errors are somewhat larger for the parameters of this third component. We 
show the spectra together with this model and also the residuals in Fig. 
\ref{Fig_fit}. The parameters of the model and their errors are also given in 
Table \ref{Tbl_fits}. 

An alternative model for the hard component is based on optically thin thermal 
emission from an isobaric cooling flow with a maximum temperature 
\citep[mkcflow in {XSPEC},][]{MuS88}. The physical picture behind this is
that accreted matter is shocked inside an optically-thin boundary layer and then
cools. This kind of model has been already applied successfully to 
a number of symbiotic stars as well as CVs \citep[e.g.][and references 
in \S \ref{sec_hard_discussion}]{MKP03}. Again, we first fitted the hard 
component only. Assuming a distance of 2.5 kpc, we set\footnote{We used a tool 
available at http://astronomy.swin.edu.au/\~{}elenc/Calculators/redshift.php to 
calculate the corresponding redshift.} the redshift to $4.16\times10^{-7}$. We 
found a temperature range in the cooling flow of 0.38 -- 60.45 keV and a 
normalization (i.e. accretion rate in M$_\odot$ yr$^{-1}$) of 
$3.11\times10^{-11}$. The absorbing column density is $2.75\times10^{23}$ 
cm$^{-2}$. The abundances are normal, 1.08 times the solar values following 
\citet{AnG89}. The reduced $\chi^2$ of this model is 1.82 and $P_{\rm nh}$ is 
16 \%. With having these parameters frozen, we added again two additional 
components: i) a Gaussian representing the peak between 6 and 7 keV with 
parameters $E = 6.10$ keV and $\sigma = 0.25$ and ii) an absorbed soft thermal 
component. The temperature and column density are similar to those derived 
before, $k\,T \sim 0.18$ and $6.0\times10^{21}$ cm$^{-2}$. The reduced $\chi^2$ 
of this model is 0.695 and $P_{\rm nh}$ is 87 \%.

The total flux is almost identical throughout all of our models, covering the 
range (1.74--1.79)$\times10^{-13}$ ergs\,cm$^{-2}$\,s$^{-1}$.

Although these two kinds of model are commonly used for the hard
component seen in symbiotic stars, we also fitted other models to the hard 
component, namely i) a highly absorbed blackbody 
($n_{\rm H} = 5.87\times10^{23}$ cm$^{-2}$, $k\,T \sim 1.11$, 
$\chi^2 = 0.9251$, $P_{\rm nh}$ is 57 \%), ii) a highly absorbed 
power-law ($n_{\rm H} = 8.02\times10^{23}$ cm$^{-2}$, $\Gamma = 4.65$,  
$\chi^2 = 0.9235$, $P_{\rm nh}$ is 58 \%). All of them give acceptable fits.

\begin{table*}
\caption{Spectral results in the 0.1--15 keV energy range for the EPIC 
  spectra. Errors are quoted at the 90 \% confidence level for a single 
  interesting parameter.}
\label{Tbl_fits}
\begin{tabular}{l|ll|lll}
  \hline\hline
  Model$^{\mathrm{a}}$ & Parameter & Unit & Value & \multicolumn{2}{l}{Error} \\
  \hline
  \multicolumn{5}{c}{Thermal model} \\
  \hline
{AP2G} & 
$n_{\rm H}$ & $10^{22}$ $\rm cm^{-2}$ & $27.87$ & ${-5.64}$ & ${7.13}$  \\
$\chi_r^2 = 0.628$ (d.o.f. = 26) & 
$k\,T$ & keV & $11.26$ & ${-3.44}$ & ${10.59}$ \\
& norm & & $3.01\times10^{-4}$ & ${-5.98}\times10^{-5}$ & ${7.46}\times10^{-5}$ 
\\
& & \\
& $E$ (GS) & keV & $6.08$ & $-0.28$ & $0.18$ \\
& $\sigma$ (GS) & keV & $0.23$ & $-0.23$ & $0.28$ \\
& norm (GS) & & $2.74\times10^{-6}$ & ${-2.74}\times10^{-6}$ & 
${1.63}\times10^{-6}$\\
& & \\
& $n_{\rm H}$ & $10^{22}$ $\rm cm^{-2}$ & $0.61$ & $-0.38$ & $0.59$ \\
& $k\,T$ & keV & $0.18$ & $-0.18$ & $5.94$ \\
& norm & & $2.17\times10^{-5}$ & ${-2.02}\times10^{-5}$ & $0.47$ \\
& Total flux & ergs\,cm$^{-2}$\,s$^{-1}$ & $1.744\times10^{-13}$ & & \\   
\hline
\multicolumn{5}{c}{Cooling flow model} \\
\hline
{CF5} & 
$n_{\rm H}$ & $10^{22}$ $\rm cm^{-2}$ & $27.51$ & $-2.88$ & $3.46$ \\
$\chi_r^2 = 0.695$ (d.o.f. = 26) & 
$T_{\rm min}$ & keV & $0.38$ & $-0.37$ & $5.51$ \\
&$T_{\rm max}$ & keV & $60.45$ & $-7.94$ & $8.23$ \\
& Abundance & & 1.086 & $-0.85$ & $1.21$ \\
& norm & & $3.11\times10^{-11}$ & $-3.46\times10^{-12}$ & $3.46\times10^{-12}$ \\
& & \\
& $E$ (GS) & keV & $6.10$ & $-0.30$ & $0.22$ \\
& $\sigma$ (GS) & keV & $0.25$ & $-0.25$ & $0.32$ \\
& norm (GS) & & $2.62\times10^{-6}$ & $-2.62\times10^{-6}$ & 
$1.71\times10^{-6}$ \\
& & \\
& $n_{\rm H}$ & $10^{22}$ $\rm cm^{-2}$ & $0.60$ & $-0.13$ & $0.59$ \\
& $k\,T$ & keV & $0.18$ & $-0.17$ & $0.05$ \\
& norm & & $2.20\times10^{-5}$ & $-2.2\times10^{-5}$ & $2.3\times10^{-5}$ \\
& Total flux & ergs\,cm$^{-2}$\,s$^{-1}$ & $1.792\times10^{-13}$ & & \\   
\hline
\end{tabular}
\medskip
\begin{list}{}{}
\item[$^{\mathrm{a}}$] Models applied in {XSPEC} notation:\\
\mbox{{AP2G = wabs(apec)+wabs(apec+gaussian) (Abundance = 1, redshift = 
0)}} \\
\mbox{{CF5 = wabs(apec)+wabs(gaussian+mkcflow) (Redshift = 
$4.16\times10^{-7}$, switch = 0)}} \\
\end{list}
\end{table*}

\subsection{Light curves}

\begin{figure}
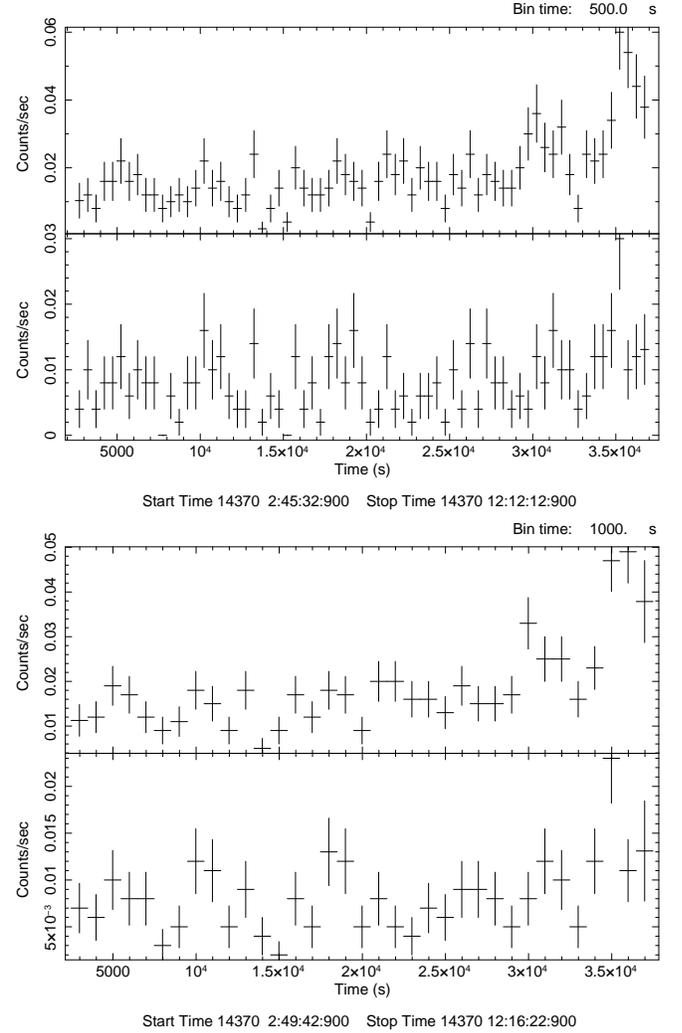

  \centering
  \includegraphics[angle=-90,width=\columnwidth]{fig3a.ps}
  \includegraphics[angle=-90,width=\columnwidth]{fig3b.ps} \\
  \caption{PN light curves binned at 500s and 1000s for energies between 0.3--15
    keV (top) and between 6--15 keV (bottom).}
  \label{Fig_lightcurve}
\end{figure}

We examined time series binned at 500 and 1000s (Fig. \ref{Fig_lightcurve}).
By eye, there seems to be quasi-periodic variation on timescales of hours. The 
ratio of measured  to expected variance is 2.781 in the 500s-binned light curve 
and 4.605 in the 1000s-binned time series. Hence this variability has a large 
statistical significance.

Within the final 4 ks of our observations, the flux increases by a factor of 
2.5--4. This increase is only pronounced at energies lower than 6 keV. 

\begin{figure}
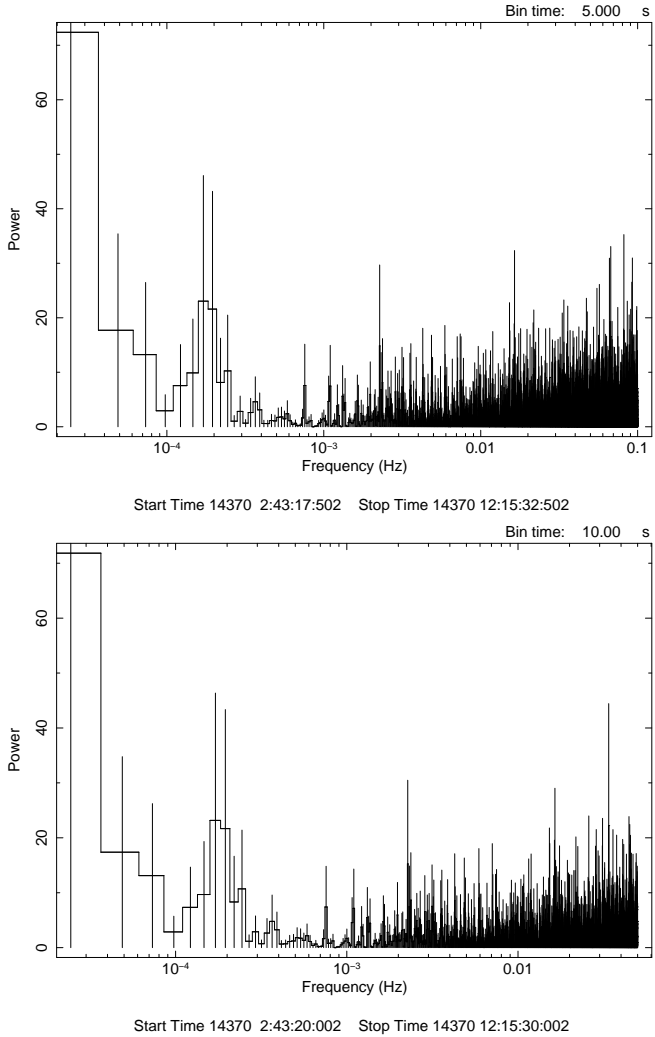

  \centering
  \includegraphics[angle=-90,width=\columnwidth]{fig4a.ps}
  \includegraphics[angle=-90,width=\columnwidth]{fig4b.ps} \\
  \caption{Power spectra of the pn X-ray light curve of MWC\,560, binned at 5 
    and 10 $s$, respectively.}
  \label{Fig_powspec}
\end{figure}

We created power spectra of the full X-ray light curve of MWC\,560, binned 
at intervals between 5 and 10 s (Fig. \ref{Fig_powspec}). We found 
several frequencies, however, the statistical significance of these peaks is 
not very high. The corresponding periods are in the range of stochastic 
flickering seen in MWC\,560 in UV and optical bands 
\citep[e.g.][]{MPS93, TKI96}. 

No hints for a periodic rapid X-ray variability are present. The fractional 
amplitude to which we are theoretically sensitive can be calculated using the
total number of counts (624) and the number (6868) of frequencies searched 
\citep[e.g.][]{LuS07}. We find that $A \sim 0.31$, i.e. we are sensitive to 
oscillations with fractional amplitudes of 31 \% in regions of the power 
spectrum dominated by white noise.

\section{Discussion} \label{sec_dis}

\subsection{The hard component of the spectrum} \label{sec_hard_discussion}

The first model fits show a highly absorbed hard ($k\,T \sim$ 11.3--13.2 keV) 
thermal component. The value for the column density of absorbing material, 
$2.8\times10^{23}$ cm$^{-2}$, is very high. Models of visual interstellar
absorption towards MWC\,560 give much lower values: $n_H = 1.5\times10^{21}$ 
cm$^{-2}$ \citep{HMS97}, $n_H = 3.07\times10^{21}$ cm$^{-2}$ \citep{KBH05} or 
$n_H = 3.25\times10^{21}$ cm$^{-2}$ \citep{DiL90}. Thus the absorption must be 
caused by material close to the central white dwarf. Our own calculations 
showed that the column density of the jet is only of the order of $10^{22}$ 
cm$^{-2}$ \citep{StC05,SKC01} or below, thus the high column 
density derived from our data cannot be attributed to the jet.

MWC\,560 seems to well fit in the class of symbiotic stars with a hard X-ray
component. Both the amount of absorbing material and the temperature of the 
hard component we found in MWC\,560 using a thermal model are close to values 
seen in similar objects as CH\,Cyg \citep[$n_H = 3.2\times10^{23}$ cm$^{-2}$, 
10 keV;][]{MIK07}, SS7317 \citep[$n_H = 1.6\times10^{23}$ cm$^{-2}$, 9.3 
keV;][]{SMM08} and R\,Aqr \citep[$n_H = 3.5\times10^{23}$ cm$^{-2}$, 
6.8 keV;][]{NDK07}. We also fit the spectra with a cooling flow model and 
find a maximum temperature of 61 keV in MWC\,560 which is similar to those in 
RT\,Cru \citep[80 keV;][]{LuS07} and T\,CrB \citep[57 keV;][]{LSM07}. This may 
be a hint that the masses of the white dwarf in these objects are comparable, 
since the maximum temperature is determined by the depth of the potential well 
around the white dwarf and thus by its compactness.

We infer from our X-ray data that in MWC\,560 the accretion rate is 
$3.1\times10^{-11}\,( d / 2.5 \textrm{kpc} )^2$ M$_\odot$ yr$^{-1}$, 
significantly lower than in T\,CrB \citep[$4.2\times10^{-9}$ M$_\odot$ 
yr$^{-1}$,][]{LSM07} and RT\,Cru \citep[$1.8\times10^{-9}$ M$_\odot$ 
yr$^{-1}$,][]{LuS07}, two objects without jets, and also much lower than in 
CH\,Cyg, which does have jets. With a mean flux of the hard energy component of 
about $1.6\times10^{-13}$ ergs\,cm$^{-2}$\,s$^{-1}$, MWC\,560 has an about 10 
times lower flux than the hard X-ray component of CH\,Cyg during the faint state
in 2006 and even an about 420 times lower flux during CH\,Cyg's bright state in 
1994 \citep{MIK07}. Since CH\,Cyg is about ten times closer than MWC\,560, a 
factor of 100 would be expected, if both objects had the same accretion rate.

We hypothesise that MWC\,560's accretion rate is variable and that MWC\,560 
switches between faint and bright states as seen in e.g. neutron star X-ray 
binaries and also CH\,Cyg. This hypothesis is supported by \citet{SKC01}'s 
determination of an accretion rate of $5\times10^{-7}$ M$_\odot$ yr$^{-1}$ from 
ultraviolet data. This high rate is consistent with a jet outflow rate of about 
$7\times10^{-9}$ M$_\odot$ yr$^{-1}$ at the time of their spectroscopic 
observations, assuming that a few percent of the accreted material is being 
ejected in the jet.

In our fits, we added a Gaussian representing an emission line. 
We found a central energy of $\sim 6.1$ keV and a width of $\sim 0.25$ keV. 
Usually, the following lines are detected: a Fe K$\alpha$ fluorescence line at 
6.4 keV or Fe XXV and XXVI emission lines at 6.7 and 6.97 keV, respectively.
Only the first of these lines is consistent with our fits, since the error of 
the peak energy is large, 0.3 keV. However, the large errors of
the line strength (see "norm (GS)", Table \ref{Tbl_fits}, both models) found in 
our fits suggest that the data are also consistent with no line being present.

\subsection{The soft jet component(s)}

In \citet{StS07}, we modeled the X-ray emission which we can expect from the 
jet in MWC\,560, according to our models of \citet{Stu06}. We found that the 
synthetic jet spectrum can be described with a hot and a warm optically-thin 
plasma component. During the minima (maxima) of the X-ray emission light cycle, 
the hot component is predicted to have a temperature of $\sim$0.7 keV (1.6 
keV); the warm component has temperature values of $\sim$0.14 keV (0.33 keV). 
The total X-ray luminosity found in our best-fit simulation using the optical
data \citep[model iv';][]{Stu06} decreased with time from an initial value of 
about $5\times10^{-12}$ ergs\,cm$^{-2}$\,s$^{-1}$ directly after the emergence of
the jet. Note that we modeled only 115 days in model iv' due to computational 
constraints. The X-ray luminosity is proportional to the density of the shocked 
material and to the shock velocity. At the beginning, the density contrast as 
well as the velocity difference between the jet and the slow ambient medium are 
the highest. After some time, however, the fast wind does not plow into the 
ambient medium anymore, but into previously ejected fast wind material with 
more similar density and velocity values as the unshocked fast wind itself. At 
the same time the jet expands also in its lateral direction away from the 
symmetry axis, i.e. lateral kinetic energy becomes an important sink of energy 
and the conversion factor of input kinetic energy into X-ray luminosity 
decreases. Both factors cause the decline in the X-ray luminosity seen in the 
simulations. Superimposed on the global decline, a cycle with a period of 7-day 
and amplitudes of about 20 \% was present.

Between 0.2 keV and 3 keV, the observed flux is about $10^{-15}$ 
ergs\,cm$^{-2}$\,s$^{-1}$. Hence \citet{MWJ97}'s sensitivity was with their 
lower limit of $8\times10^{-15}$ ergs\,cm$^{-2}$\,s$^{-1}$ just below of what 
had been needed to detect MWC\,560. Therefore the highest possible decline, if 
any, of the soft X-ray emission is about a factor of eight within the 15 years 
from the ROSAT observations in April 1992 and our observations in September 
2007. The observations of \citet{MeB43}, where blue-shifted absorption features 
near the emission lines of HI and Ca II K were already noticed, set a lower 
limit on the age of the jet of about 60 yrs.

\begin{figure*}
  \centering
  \includegraphics[width=\textwidth]{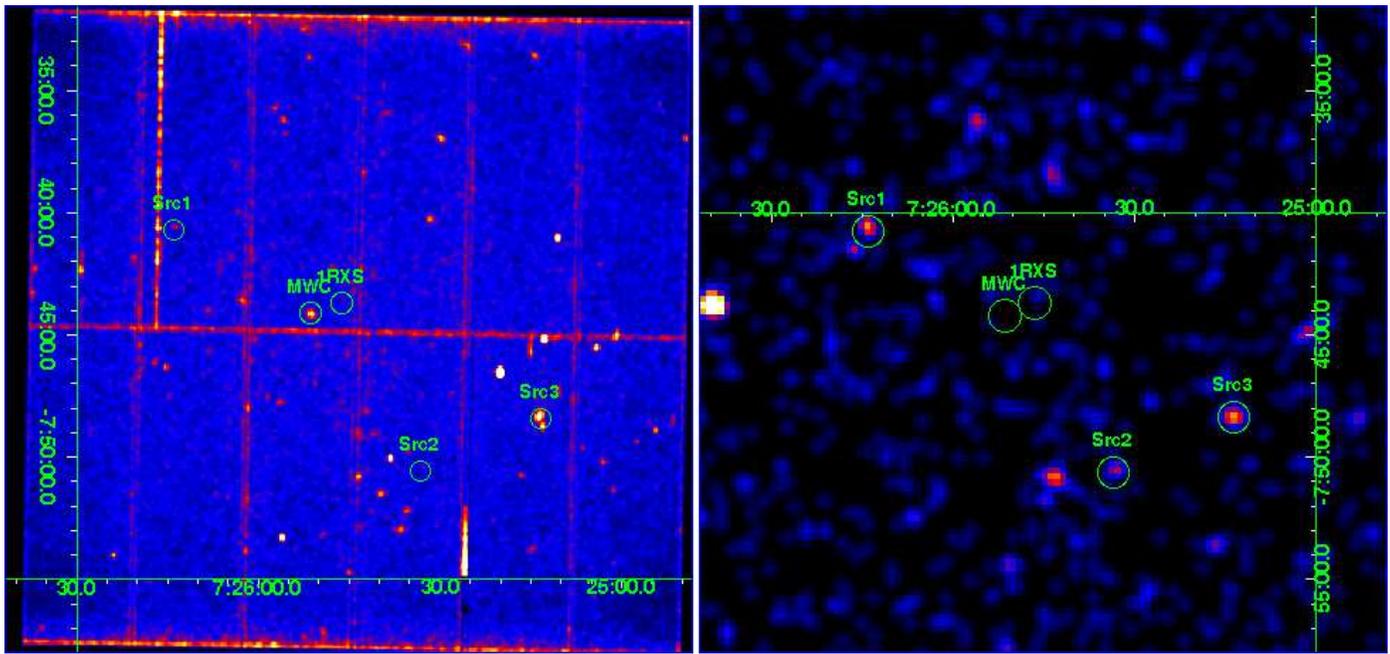}
  \caption{Images from our full XMM pn field of view (left) and the 
    same field of view within the ROSAT observation of \citet{MWJ97} (right). 
    Also plotted are the three closest sources listed in the ROSAT All-Sky 
    Survey Faint Sources catalog (Src1: 1WGA J0726.2-0740, Src2: 
    1WGA J0725.5-0750, Src3: 1WGA J0725.2-0748). For all these source, we can 
    find counterparts at the correct position in the ROSAT and XMM 
    observations. The source 1RXS is not present in either 
    of them. MWC\,560 has not been detected by ROSAT.}
  \label{Fig_rosat_xmm}
\end{figure*}

The ROSAT All-Sky Survey Faint Sources catalog lists a source, 
1RXS\,J072546.4-074342, 1RXS hereafter, which lies 77'' away from the 
SIMBAD position of MWC\,560. Its count rate in October 1990 was 0.0149 counts 
s$^{-1}$ (a flux of $1.7\times10^{-13}$ ergs\,cm$^{-2}$\,s$^{-1}$), i.e. the flux
is 100 times higher than our observed flux. In Fig. \ref{Fig_rosat_xmm}, we plot
our full XMM pn field of view (left) and the same field of view within the ROSAT
observation of \citet{MWJ97} in April 1992 (right). We also identified the 
three closest sources listed in the ROSAT All-Sky Survey Faint Sources catalog. 
For all these source, we can find counterparts at the correct position in the 
ROSAT and XMM observations. The source 1RXS is not present in either of them. 
Thus, 1RXS has a large and variable X-ray flux, highlighting its 
potential importance as an X-ray transient source. However, due to the absence 
of obvious pointing errors between XMM and ROSAT data, we conclude 
that 1RXS is not the same source as MWC560.

The observed temperature of the soft component, $k\,T \sim 0.18$, is slightly 
lower than predicted by our simulations. The observed flux values between 0.2 
keV and 3 keV are more than a factor 100 less than predicted by our models 
($1.6 \times 10^{-13}$ erg s$^{-1}$ cm$^{-2}$ between 0.1 -- 2.4 keV), and 
clearly cannot be due to coincidentally catching the MWC560 jet at the minimum 
of the 7-day flash cycle. Extrapolating the global decline in our models, 
however, shows that the flux from the jet can drop to the observed fluxes 
within several years after the emergence of the jet, if we assume that the jet 
emission follows the same ejection pattern as in our models.

Although we know that a jet is present in MWC\,560 and requiring only the jet 
for explaining the observed soft component is tempting, there are other 
scenarios for producing soft or even supersoft X-rays. A component with a 
temperature of about 0.2 keV is also typical for photospheric emission from 
very hot white dwarfs \citep[e.g.][]{JMW94}. Soft components which peak around 
0.8--1 keV are produced by several kinds of shocks with shock velocities of 
about 1000 km\,s$^{-1}$, present not only in jets, but also in colliding winds 
and during the accretion process.

We note that our soft component is resolved in only three energy bins, since
we required at least 25 counts per bin. This makes it difficult to fit this 
component in a reliable manner and so the resulting constraints on our jet 
model are very weak or non-existent. Furthermore it is not possible to 
distinguish between the above mentioned scenarios producing supersoft or soft 
X-ray emission. We have to obtain more data by observing MWC\,560 again with 
either Chandra or XMM.

\subsection{The timing analysis}

We detected quasi-periodic flickering on timescales of minutes and hours, which
usually emanates from an accretion region close to the white dwarf. Accretion 
thus seems to power the detected X-ray emission. One open question, however, is 
whether the white dwarf is magnetized or not. In the first case, the hard 
component is emitted by accretion columns. The disk will be truncated, the BL 
disappears and matter falls freely along magnetic field lines towards the 
magnetic poles. The infalling matter then forms a strong shock near the white 
dwarf surface leading to optically-thin hard X-ray emission with $kT \sim$ 
10 -- 20 keV  \citep[e.g.][]{CWR00}. Also the presence of rapid {\em periodic} 
variability with the Keplerian period at the truncation radius would be very 
likely. If the white dwarf is not magnetized , the hard component is emitted by 
an optically-thin boundary layer \citep{PrS79, PoN95}. Since we did not detect 
periodic variability, we favor this possibility.

The increase seen at the end of our previous observation may (i) be related to 
the emergence of a new high-velocity pulse creating an X-ray flash as seen in 
our simulations, (ii) represent a secular increase, or (iii) be due to 
switching to a ``bright" state from a ``faint" one, as seen in CH\,Cyg 
-- only new observations can distinguish between these alternatives.

\section{Conclusion}

We discovered X-ray emission from MWC\,560 using XMM-Newton. We found spectra 
consisting of two distinct components, a soft and a hard component. Thus 
MWC\,560 is the newest member of a new sub-class of symbiotic stars emitting 
hard X-rays with similar parameters. 

From modelling the hard component, the absence of periodic variability and the 
presence of short-term flickering, we conclude that the hard component is 
emitted by an optically-thin boundary layer around a high-mass white dwarf. 

We hypothesise that MWC\,560's accretion rate is variable and that MWC\,560 
switches between faint and bright states as seen in e.g. neutron star X-ray 
binaries and also CH\,Cyg. 

Due to low photon counts in the soft component, it is difficult to fit this 
component in a reliable manner. Furthermore it is not possible to 
distinguish between several scenarios producing supersoft or soft X-ray 
emission. On the other hand, the soft X-ray flux, significantly smaller 
than predicted by our models at the emergence of the jet, still provided 
a strong constraint on our simulations, testing of which was the main goal of 
the presented observations.

More observational data with either XMM or Chandra are needed as well as 
further simulations with longer time baselines.

\acknowledgements
We acknowledge helpful and improving comments and suggestions by the referee. 
The present work was supported in part by the European Community's Marie
Curie Actions - Human Resource and Mobility within the JETSET (Jet
Simulations, Experiments and Theory) network under contract
MRTN-CT-2004 005592 (MS). RS thanks NASA for funding this work by XMM-Newton 
AO-6 award NMO710766-103905, an LTSA award (NMO710840-102898); RS also received 
partial support for this work from an HST GO award (no. GO-10317.01) from the 
Space Telescope Science Institute (operated by the Association of Universities 
for Research in Astronomy, Inc., under NASA contract NAS5-26555). Some of the 
research described in this paper was carried out by RS at the Jet Propulsion 
Laboratory, California Institute of Technology, under a contract with the 
National Aeronautics and Space Administration.

\end{document}